\documentclass[
prd,
rsi,%
 amsmath,amssymb,
 reprint,%
]{revtex4-1}

\usepackage{graphicx}
\usepackage{dcolumn}
\usepackage{bm}
\newcommand{\eh}{$e^-h^+$}
\usepackage[mathlines]{lineno}

\usepackage[normalem]{ulem} 
\usepackage{xcolor}

\begin{document}


\title{Measuring the Impact Ionization and Charge Trapping Probabilities in SuperCDMS HVeV Phonon Sensing Detectors}

\affiliation{Department of Physics, Stanford University, Stanford, CA 94305 USA}
\affiliation{Department of Physics, University of California, Berkeley, CA 94720 USA}
\affiliation{SLAC National Accelerator Laboratory/Kavli Institute for Particle Astrophysics and Cosmology, Menlo Park, CA 94025 USA}
\affiliation{Fermi National Accelerator Laboratory, Batavia, IL 60510, USA}
\affiliation{Department of Physics, Santa Clara University, Santa Clara, CA 95053 USA}

\author{F.~Ponce} \affiliation{Department of Physics, Stanford University, Stanford, CA 94305 USA}
\author{W.~Page} \affiliation{Department of Physics, University of California, Berkeley, CA 94720 USA}
\author{P.L.~Brink} \affiliation{SLAC National Accelerator Laboratory/Kavli Institute for Particle Astrophysics and Cosmology, Menlo Park, CA 94025 USA}
\author{B.~Cabrera} \affiliation{Department of Physics, Stanford University, Stanford, CA 94305 USA} \affiliation{SLAC National Accelerator Laboratory/Kavli Institute for Particle Astrophysics and Cosmology, Menlo Park, CA 94025 USA}
\author{M.~Cherry} \affiliation{SLAC National Accelerator Laboratory/Kavli Institute for Particle Astrophysics and Cosmology, Menlo Park, CA 94025 USA}
\author{C.~Fink} \affiliation{Department of Physics, University of California, Berkeley, CA 94720 USA}
\author{N.~Kurinsky} \affiliation{Fermi National Accelerator Laboratory, Batavia, IL 60510, USA}
\author{R.~Partridge} \affiliation{SLAC National Accelerator Laboratory/Kavli Institute for Particle Astrophysics and Cosmology, Menlo Park, CA 94025 USA}
\author{M.~Pyle} \affiliation{Department of Physics, University of California, Berkeley, CA 94720 USA}
\author{B.~Sadoulet} \affiliation{Department of Physics, University of California, Berkeley, CA 94720 USA}
\author{B.~Serfass} \affiliation{Department of Physics, University of California, Berkeley, CA 94720 USA}
\author{C.~Stanford} \affiliation{Department of Physics, Stanford University, Stanford, CA 94305 USA}
\author{S.L.~Watkins} \affiliation{Department of Physics, University of California, Berkeley, CA 94720 USA}
\author{S.~Yellin} \affiliation{Department of Physics, Stanford University, Stanford, CA 94305 USA}
\author{B.A.~Young} \affiliation{Department of Physics, Santa Clara University, Santa Clara, CA 95053 USA}

\date{\today}

\begin{abstract}
A 0.93 gram $1{\times}1{\times}0.4$~cm$^3$ SuperCDMS silicon HVeV detector operated at 30 mK was illuminated by 1.91~eV photons using a room temperature pulsed laser coupled to the cryostat via fiber optic. The detector's response under a variety of specific operating conditions was used to study the detector leakage current, charge trapping and impact ionization in the high-purity Si substrate. The measured probabilities for a charge carrier in the detector to undergo charge trapping (0.713~$\pm$~0.093\%) or cause impact ionization (1.576~$\pm$~0.110\%) were found to be nearly independent of bias polarity and charge-carrier type (electron or hole) for substrate biases of $\pm$~140~V.

\end{abstract}



\maketitle

The lack of evidence of supersymmetry at the LHC has spurred additional interest in light dark matter (DM) candidates such as axions, dark photons, and other hidden sector entities~\cite{Essig2012, LightDarkSectors, DarkSectors, Nelson2011, Holdom1986}. The search for these hypothesized interactions requires detectors with sub-eV energy resolution and threshold, which has motivated R\&D efforts to build detectors with single charge detection capabilities~\cite{Romani2018, Tiffenberg_17prl_Sensei}. Using these detectors to set new DM constraints or to make a discovery requires accurate detector models and simulations. These models and simulations must include the detector properties (crystal orientation, intrinsic purity, operating conditions, etc.), as well as the effects of known backgrounds (radioactivity, leakage current, etc.).

Recently developed SuperCDMS HVeV detectors provide the sensitivity necessary for modern experiments to search for light dark matter. The HVeV detector makes use of the Neganov-Trofimov-Luke (NTL) effect~\cite{Neganov1985, Luke1988} by applying a bias voltage between opposite faces of a high-purity Si substrate. This voltage biasing scheme converts ionization energy created by a single event into an amplified phonon signal that is then read out using superconducting sensors on one face of the detector. 

Early experiments showed that HVeV detectors provide charge quantized output signals when illuminated with 1.91 eV photons~\cite{Romani2018}. While the observed event histogram peaks corresponding to integer numbers of \eh\ pairs detected were Gaussian, sub-gap infrared photons (SGIR) added significant ``fill-in'' between the quantized peaks. The same detector was later run with an improved fiber optic setup and IR-absorbing windows that confirmed the initial SGIR hypothesis~\cite{CDMS2018_DMSearch}. But even with the improved optical system there remained an estimated 3\% ``fill-in'' between quantized energy peaks that was attributed to a combination of charge trapping and impact ionization in the Si substrate. Charge trapping occurs when, e.g., an  electron (or hole) falls into a vacancy and gets stuck; this reduces the total number of event related electrons (or holes) traversing the crystal, leading to low energy tails on the histogram peaks. Impact ionization occurs when a charge moving through the crystal has sufficient energy to liberate an additional charge that is loosely bound in the crystal; this process increases the total number of charges traversing the detector and produces high energy tails on the histogram peaks. This paper describes experiments performed with this detector to study charge leakage, charge trapping and impact ionization probabilities for HVeV detectors based on recently developed first-order models~\cite{Ponce2019}.

The experiments described below used a SuperCDMS silicon HVeV detector. The detector consists of a $1{\times}1{\times}0.4$~cm$^3$ high-purity Si crystal (0.93~g) patterned with quasiparticle-trap-assisted electro-thermal-feedback transition-edge sensors (QETs), and an Al parquet pattern \cite{Romani2018}. The detector was cooled to 30~mK in a dilution refrigerator and the QET sensors were voltage biased at $\sim$22\% of their normal state resistance. The bias conditions corresponded to a sensor bias power of 0.17 pW for stable operation within the tungsten TES superconducting-to-normal transition.


A single mode fiber optic was used to illuminate the Al parquet side of the detector with 650~nm (1.91~eV) photons from a pulsed laser at an adjustable repetition rate. Coarse control of the laser intensity at the detector was achieved using combinations of external optical attenuators (OA) at room temperature. Fine control of the intensity was achieved by changing the laser output power and pulse width. 

The HVeV Si substrate was ``neutralized'' at the start of the experiment by grounding the metal films on both sides of the detector (QET sensors and Al parquet) and pulsing the laser at 200~Hz with a relatively high intensity ($\sim$3$\times$10$^{16}$ photons per pulse) for ~16 hours. Physics data were collected using a fixed laser pulse width of 200~ns, -80~dB OA and a combination of two Si crystal bias voltages: $\pm$140~V, and four laser intensities: ``zero'' (no photons, 0.5~Hz, 20~$\mu$W), ``high'' ($\sim$0.5 photons per pulse, 200~Hz, 2000~$\mu$W), ``medium'' ($\sim$0.05 photons per pulse, 200~Hz, 200~$\mu$W), and ``low'' ($\sim$0.025 photons per pulse, 2000~Hz, 20~$\mu$W) for a total of eight configurations. At each of the two Si crystal biases used, the laser intensity was cycled in a specific order and time distribution, given by: 9.1\% zero, 30.3 \% high, 30.3\% medium, and 30.3\% low intensity. Prior to each acquisition (data collected using a single configuration during one cycle), the Si crystal was pre-biased at +($-$)160~V for one minute followed by reducing the crystal bias to +($-$)140~V for one minute. 

Data were recorded in a semi-continuous mode at a sample rate of 625 kHz, using a trace length of ~1.68~sec (2$^{20}$ samples) triggered by the internal TTL of the laser. We purposefully discarded the one laser-induced event in each ``zero" intensity trace. A total live-time of 15.4 (9.6) hours before cuts was collected at a detector polarity of +($-$)140~V over 27 ($<$18) hours of real-time. 

An aggressive raw-time cut was applied to remove all traces that contained a high-energy event. This was needed to avoid processing real signals that ride on the tail of a high energy pulse or that get distorted in electronics because of a DC voltage baseline shift in the QET readout caused by the energetic event. The raw-time cut reduced the total live-time by $\sim$70-75\%. 

\begin{figure}[ht!]
    \begin{center}
    \includegraphics[height=3in]{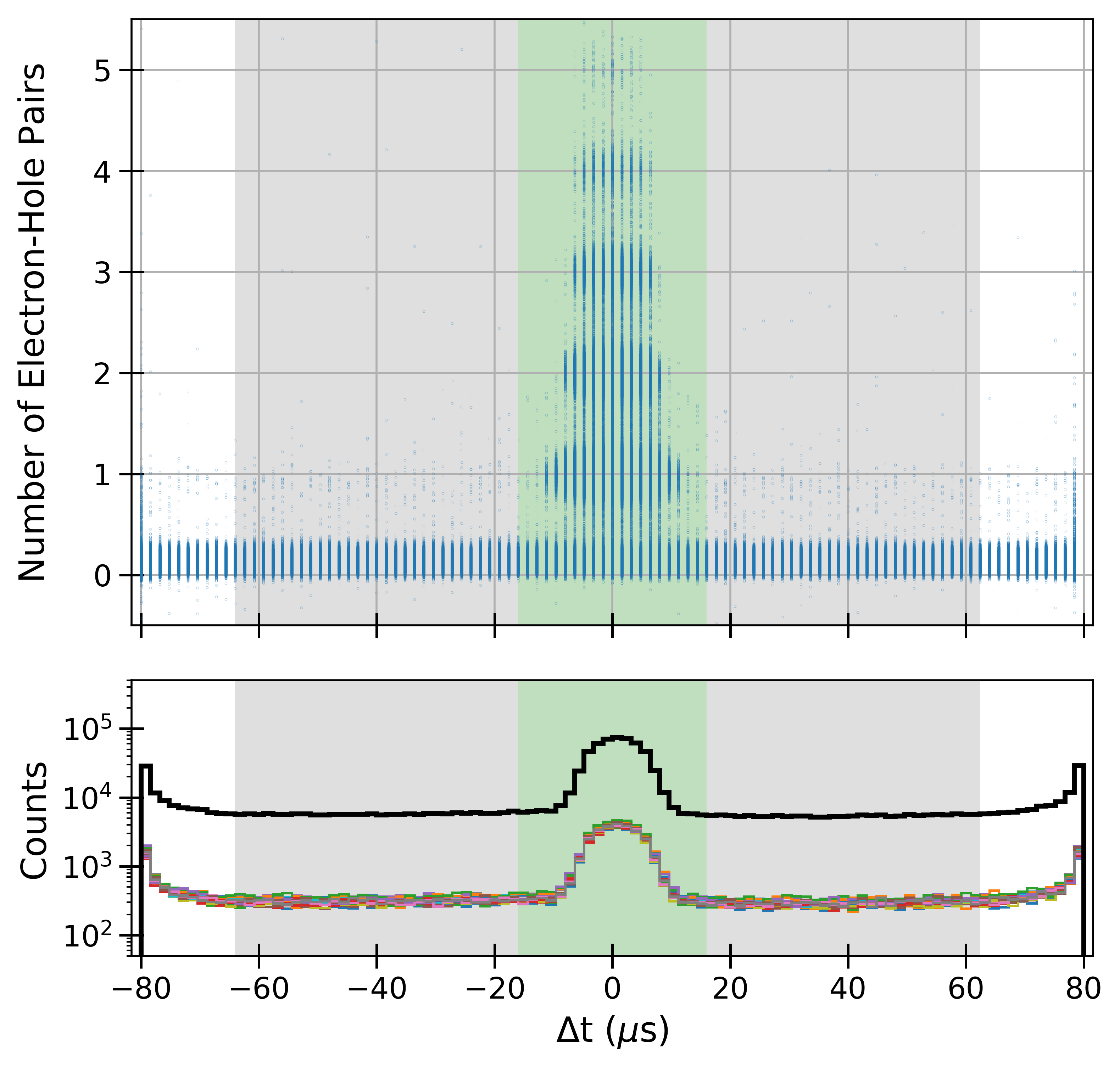}
    \caption{\footnotesize (color online) (\textit{Top}) Scatter plot of event arrival times relative to laser pulse trigger. Events in which photons from the laser were absorbed show up in a cluster (green highlight). Events outside this range correspond to laser pulses where no photons were absorbed in the detector. The gray regions mark the events used to study the leakage rates in the background. (\textit{Bottom}) Histogram of the \textit{top} scatter plot showing how the first and last 16~$\mu$s have edge effects due to the search window. The non-highlighted region was excluded in this analysis.}
    \label{Arrival_Time_Histogram}
    \end{center}
\end{figure}

An optimal filter (OF) was generated from a 1~ms pulse template and noise PSD derived from each acquisition. The OF was inverse Fourier transformed to carry out the analysis in the time domain by convolving the transformed OF with the full trace to get an OF amplitude as a function of time. The laser TTL signal was used to identify “laser events”. We associated the largest amplitude pulse within $\pm$~80~$\mu$s centered on the laser TTL trigger as the time-shifted OF amplitude and the corresponding position as the relative arrival time for the ``laser event'' (regardless of whether a true energy deposition occurs within that time period).  Pulse pile-up was removed by applying a flat $\chi^2$ cut, which had a passing fraction of 99\% at the quantized laser peaks. 

There was a slight drift in detector gain of $\sim$ $-$5\% over the course of 27 hours of real-time for the +140~V crystal bias data. The detector stability over long periods of time enabled us to use the high-intensity laser data sets to calibrate all data sets in the same cycle: zero, high, medium, low. A quadratic calibration of the form $ax(1 + bx)$ was performed using the centroids from Gaussian fits to the 1, 2, and 3 \eh\ pair peaks. The non-linearity, b, was on the order of 3\%, which was consistent with prior measurements using more peaks at higher intensity~\cite{CDMS2018_DMSearch}.



\begin{figure}[ht!]
    \begin{center}
    \includegraphics[height=3in]{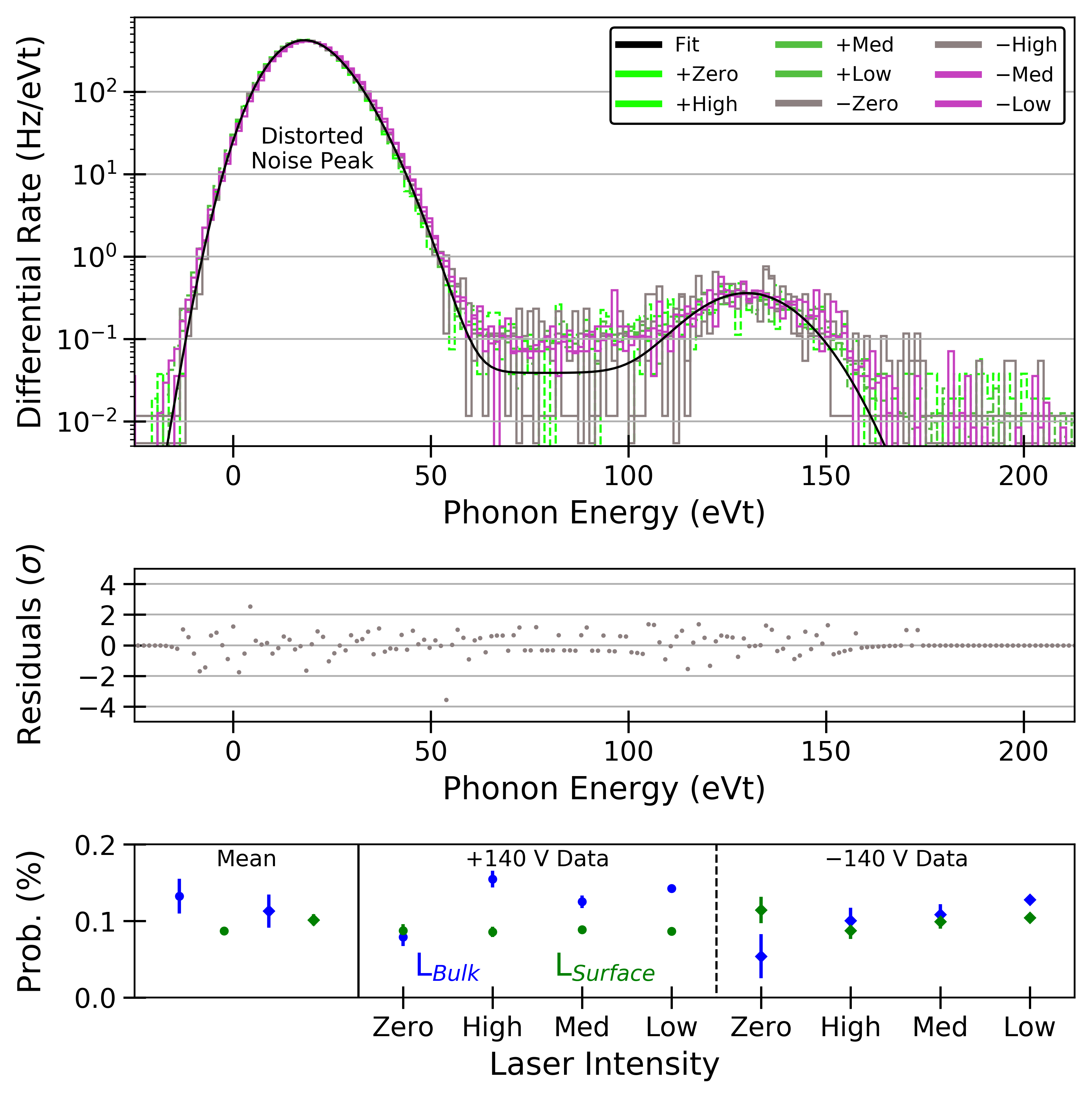}
    \caption{\footnotesize (color online) (\textit{Top}) Background spectra (multi-colored lines) for the eight configurations and the fit for the high laser intensity with $-$140~V substrate bias. The spectra were normalized by the reduced total live-time. (\textit{Middle}) Residuals for the fit normalized by the counting statistics of each bin. Bins with zero counts were artificially set to zero. (\textit{Bottom}) The measured bulk (blue) and surface (green) leakage probabilities at +140~V (circles) and $-$140~V (diamonds) are shown to the right of the solid line; the corresponding weighted averages and standard deviations are shown to the left of the line.}
    \label{bkgd}
    \end{center}
\end{figure}

Figure~\ref{Arrival_Time_Histogram} (top) shows the scatter plot of calibrated time-shifting OF amplitudes versus relative arrival times for the +140~V bias high-intensity data. Events where laser photons were absorbed cluster between $-$16 and 16~$\mu$s (green shade). Only noise/leakage events appeared outside the green shaded region. The sudden increase in noise/leakage events in the first and last 16~$\mu$s of the 160 ~$\mu$s-wide window (Figure~\ref{Arrival_Time_Histogram}, bottom) were attributed to leakage events outside the search window. These events were discarded from the main analysis. This cut disproportionately affects 0~\eh\ pair event statistics, which was accounted for by adding a fit parameter to the 0~\eh\ pair amplitude. The events in the gray region of Figure~\ref{Arrival_Time_Histogram} were used to generate the corresponding background spectrum for each configuration. Events in the combined (green + gray) shaded regions (i.e., a 128~$\mu$s search window) were used to determine the impact ionization and charge trapping probabilities for this detector.

We model our leakage current background, $B(x)$, as a noise peak with a continuous distribution of bulk leakage and quantized surface leakage~\cite{Ponce2019}:
\begin{eqnarray}
    & &B(x) = 
    \frac{\textrm{L}_0\textrm{N}e^{-\frac{(x - c_0)^2}{2\sigma^2}}}{\sqrt{2\pi\sigma^2}}\left(\frac{\left(1 - erf\left(\frac{x - c_0}{\sqrt{2\sigma^2}}\right)\right)}{2}\right)^{N-1}\nonumber\\
    & &+ \frac{\textrm{L}_{\textrm{Surf}}}{\sqrt{2\pi\sigma^2}}e^{-\frac{(x - c_1)^2}{2\sigma^2}}\nonumber\\
    & &+ \frac{\textrm{L}_{\textrm{Bulk}}}{2(c_1 - c_0)}\left(erf\left(\frac{x - c_0}{\sqrt{2\sigma^2}}\right) - erf\left(\frac{x - c_1}{\sqrt{2\sigma^2}}\right)\right)
\end{eqnarray}
where N is the effective number of independent measurements within the OF search window, $\sigma$ is the detector resolution,  L$_{\textrm{Bulk}}$ is the bulk leakage probability, L$_\textrm{{Surf}}$ is the surface leakage probability, L$_0$ = (1 - L$_\textrm{{Bulk}}$ - L$_\textrm{{Surf}}$), and $c_{0}$ ($c_{1}$) is the centroid of the quantized 0$^{th}$ (1$^{st}$) \eh\ pair peak. The inclusion of c$_0$ in the first term was due to an offset introduced by the time-shifting OF.

\begin{figure}[ht!]
    \begin{center}
    \includegraphics[height=3in]{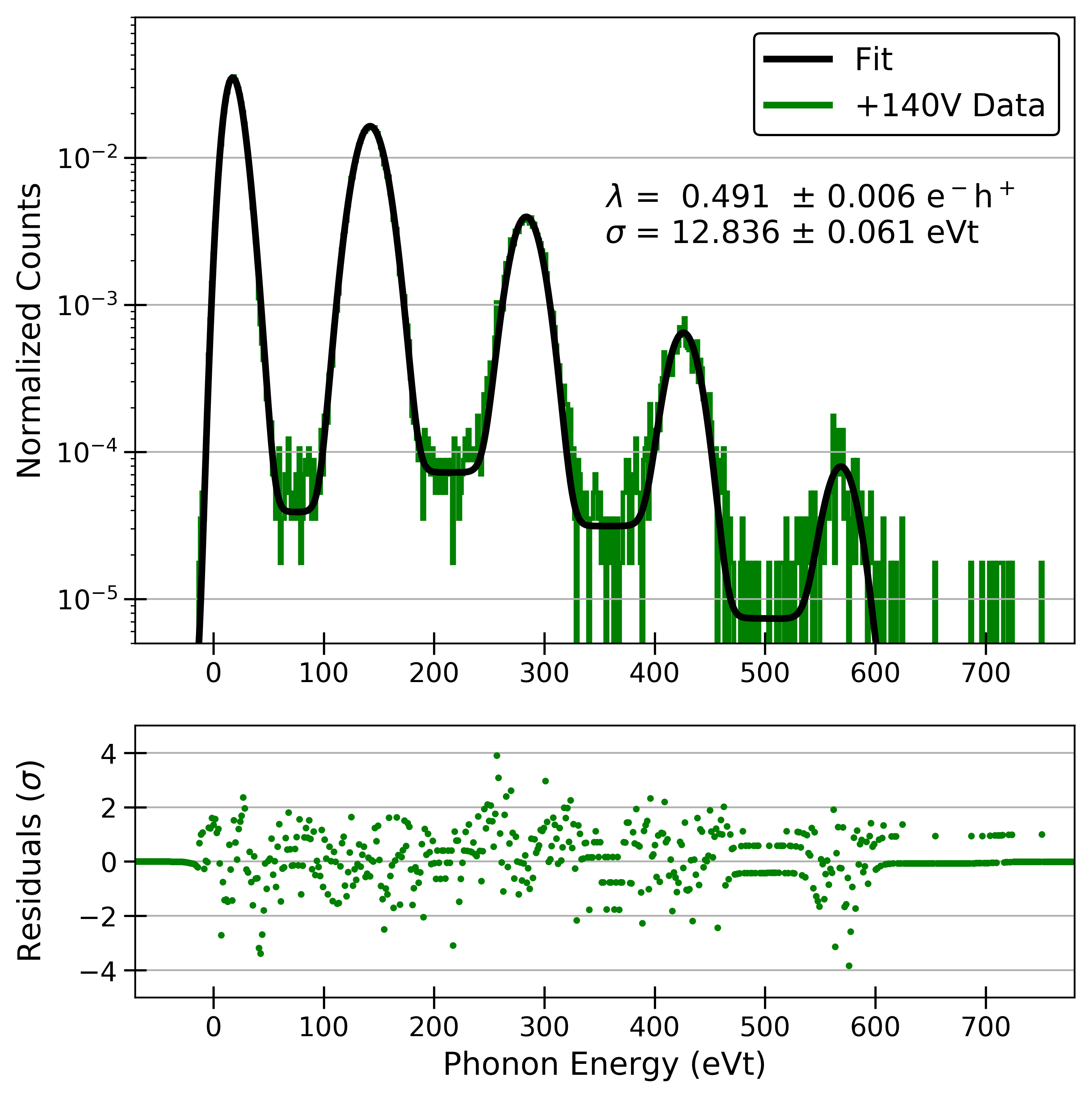}
    \caption{\footnotesize (color online) (\textit{Top}) Spectrum of laser-induced events (green) after cuts ($\sim$4 minutes), with analytical fit (black line) that includes charge leakage, charge trapping and impact ionization. 
    (\textit{Bottom}) Residuals normalized by the bin counting statistics. Bins with zero counts were artificially set to zero.}
    \label{Sample_Imp_Trap_Fitting}
    \end{center}
\end{figure}

The observed background as a function of eVt (the total phonon energy in eV produced by an event) for all eight configurations are shown in Figure~\ref{bkgd}. The spectra were normalized by the reduced total live-time (number of events times the search window length of 128~$\mu$s). No significant change in the background was observed throughout the full 48 hour period of data taking, as evidenced by the nominally identical profiles shown in Figure~\ref{bkgd} (top). Figure~\ref{bkgd} (middle) shows the residuals (gray circles) for the $-$140~V high intensity data fit (top panel, black curve), lie mostly within 2$\sigma$ of the bin uncertainty indicating a good fit to our model. Bins with zero counts were artificially set to zero. Figure \ref{bkgd} (bottom) shows the fitted bulk (blue) and surface (green) leakage probabilities for the two crystal bias polarities: +140~V (circles) and $-$140~V (diamonds). 

The bulk leakage data at $\pm$140~V varied over a narrow range with the zero intensity values significantly lower than the other fits. This discrepancy may be due to the laser TTL signal introducing electronic cross talk; however, much effort was invested to mitigate such effects and no cross talk was observed when averaging over 100 traces. We observed a weighted bulk leakage event probability (blue points, left of solid black line) of 0.132~$\pm$~0.023\% at +140~V and 0.113~$\pm$~0.022\% at $-$140~V and concluded that the bulk leakage does not depend on the crystal bias polarity. 

\begin{figure}[ht!]
    \begin{center}
    \includegraphics[height=3in]{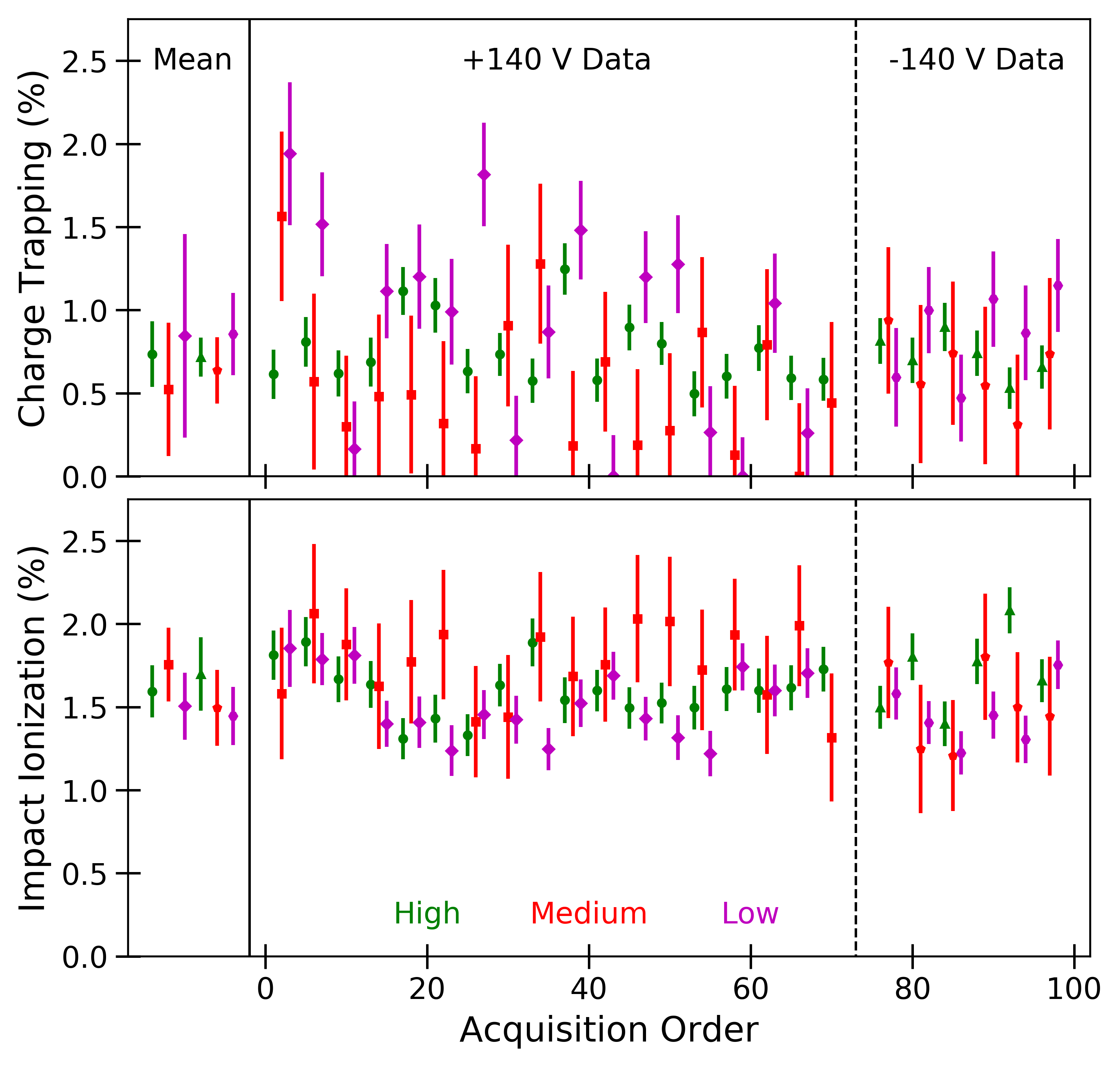}
    \caption{\footnotesize (color online) (\textit{Top}) Charge trapping and (\textit{bottom}) impact ionization probabilities for all acquisitions taken over the course of two days (left of solid black line). The weighted average and standard deviations are shown to the left of the black solid line with the individual $\pm$140~V data plotted to the right of the solid line separated by the dashed black line. Values were fitted while holding the bulk and surface leakage probabilities fixed using the background spectrum for each crystal bias and laser intensity (Figure~\ref{bkgd} bottom left of solid line)}
    \label{itparam}
    \end{center}
\end{figure}

The surface leakage data at +140~V were statistically equivalent, while the $-$140~V data varied with some overlapping uncertainties. We observed a weighted surface leakage event probability (green points, left of solid black line) of 0.087~$\pm$~0.001 for the +140~V data and 0.101~$\pm$~0.007 for the $-$140~V data. The difference indicates a very small dependence on crystal polarity although this may also be indicative of the lower statistics for the $-$140~V data. The bulk and surface leakage terms for each configuration (right side of solid line in bottom plot) were used as fixed parameters in the later fit of the impact ionization and trapping probabilities. 


We used the model outlined in Ponce et al.~\cite{Ponce2019} Equation~3 and assume the interaction of a single \eh\ pair with the crystal as having some constant probability of inducing impact ionization (effectively, generating additional charge), charge trapping (effectively removing a charge), or having the original charges move through the crystal unhindered (resulting in a quantized signal). 

In our analysis, the individual peaks $^m$h(x) were convolved with the detector Gaussian response scaled by the appropriate Poisson probabilities for the laser intensity and summed together with the background. The fitted model was
\begin{equation}
    M(x) = \kappa{}P_0(\lambda)\cdot{}B(x) + \sum_{m = 1}^{m_{max}}P_m(\lambda)(^{(m)}h\circledast{}G(\sigma))(x) 
\end{equation}
where $\kappa$ accounts for the relative arrival time cut, G($\sigma$) is the normalized Gaussian function and P$_m$($\lambda$) is the Poisson probability for peak ``m'' with an average of $\lambda$. A sample fit for a +140~V high intensity data set is shown in Figure \ref{Sample_Imp_Trap_Fitting}. The residual shows several points outside the 2$\sigma$ threshold, which may be indicative of pulse pile-up very close to the laser TTL trigger. 

A time sequence of the measured charge trapping and impact ionization probabilities for all acquisitions are shown to the right of the vertical black line in Figure~\ref{itparam}. The wide measurement distributions and large uncertainties for the medium and low laser intensity data come from the inherently poor statistics. The weighted average and standard deviations were in agreement and no dependence on the system configuration was observed. Thus, the probabilities for both holes and electrons getting across the crystal was nominally equal. Combining all the data we measure a charge trapping probability of 0.713~$\pm$~0.093\% and an impact ionization probability of 1.576~$\pm$~0.110\%.

A 0.93 gram SuperCDMS HVeV detector was operated in a semi-continuous mode and used to demonstrated the use of a time-domain OF to analyze data. Triggered pulses could be identified based on the OF estimate arrival time to within 32~$\mu$s. Data from outside this 32~$\mu$s window was used to obtain a background spectrum that was modeled to first-order as the combination of a continuous bulk and a quantized leakage currents. The model was found to be in good agreement with the full data set. A simple impact ionization and charge trapping model for a single \eh\ pair~\cite{Ponce2019} was then used to fit the detector response to six setup configurations (three non-zero laser intensities, two crystal bias polarities). By fixing the bulk and surface leakage parameters the impact ionization and charge trapping probabilities for the HVeV detector were successfully measured. 

This work was supported in part by the U.S. Department of Energy and by the National Science Foundation. This document was prepared by using resources of the Fermi National Accelerator Laboratory (Fermilab), a U.S. Department of Energy, Office of Science, HEP User Facility. Fermilab is managed by Fermi Research Alliance, LLC (FRA), acting under Contract No. DE-AC02-07CH11359. SLAC is operated under Contract No. DEAC02-76SF00515 with the U.S. Department of Energy. The authors are also especially grateful to the staff of the Varian Machine Shop at Stanford University for their assistance in machining the parts used in this experiment.

\bibliographystyle{apsrev4-1-JHEPfix-autoEtAl}
\bibliography{aipsamp}

\end{document}